\documentclass[aps,prl,twocolumn,groupedaddress,showpacs]{revtex4-1}
\pdfoutput=1
\usepackage{amsmath,graphicx,amssymb}
\usepackage{multirow,color}
\usepackage{url}
\usepackage{verbatim}
\usepackage{bm}

\renewcommand{\vec}[1]{\bm{#1}}

\newcommand{\Ceps}{C_\varepsilon}
\newcommand{\Cinf}{C_{\varepsilon,\infty}}

\newcommand{\vep}{\varepsilon}

\newcommand{\Rl}{R_{\lambda}}
\newcommand{\beq}{\begin{equation}}
\newcommand{\eeq}{\end{equation}}

\newcommand{\moritz}[1]{\textcolor{black}{#1}}

\begin{document}
\title{Nonuniversality and finite dissipation in decaying magnetohydrodynamic turbulence}

\author{M.~F. Linkmann}
\email[]{m.linkmann@ed.ac.uk}
\affiliation{SUPA, School of Physics and Astronomy, University of Edinburgh, Peter Guthrie Tait Road, EH9 3FD, UK}

\author{A. Berera}
\email[]{ab@ph.ed.ac.uk}
\affiliation{SUPA, School of Physics and Astronomy, University of Edinburgh, Peter Guthrie Tait Road, EH9 3FD, UK}

\author{W.~D. McComb}
\affiliation{SUPA, School of Physics and Astronomy, University of Edinburgh, Peter Guthrie Tait Road, EH9 3FD, UK}

\author{M.~E. McKay}
\affiliation{SUPA, School of Physics and Astronomy, University of Edinburgh, Peter Guthrie Tait Road, EH9 3FD, UK}

\begin{abstract}
A model equation for the Reynolds number dependence of the
dimensionless dissipation rate
in freely decaying homogeneous magnetohydrodynamic turbulence in
the absence of a mean magnetic field is derived from the real-space energy
balance equation, leading to
$C_{\varepsilon}=C_{\varepsilon, \infty}+C/R_- +O(1/R_-^2))$,
where $R_-$ is a generalized Reynolds number.
The constant $C_{\varepsilon, \infty}$ describes the
total energy transfer flux.
This flux depends on magnetic and cross helicities, because these affect the 
nonlinear transfer of energy, suggesting that the value of $C_{\varepsilon,\infty}$ is not universal. 
Direct numerical simulations were conducted
on up to $2048^3$ grid points, showing good agreement between data and the
model.
The model suggests that the magnitude of cosmological-scale
magnetic fields is controlled by the values of the vector field correlations. 
The ideas introduced here can be used to derive similar model equations for other turbulent
systems.
\end{abstract}

\pacs{47.65.-d, 52.30.Cv, 47.27.Jv, 47.27.Gs} % MHD (fluids), MHD (plasma physics), high Re fluids, Isotropic turb 

\maketitle

Magnetohydrodynamic (MHD) turbulence is present in many areas of physics, 
ranging from industrial applications such as liquid metal technology to 
nuclear fusion and plasma physics, geo-, astro- and solar physics, and 
even cosmology.
The numerous different MHD flow types that arise in different settings 
due to anisotropy, alignment, different values of the diffusivities, to name only a few, 
lead to the question of universality in MHD turbulence, which has been 
the subject of intensive research by many groups 
\cite{Dallas13a,Dallas13b,Wan12,Schekochihin08,Mininni11,
Grappin83,Pouquet08,Beresnyak11,Boldyrev11,Grappin10,Lee10,Servidio08}.
The behavior of the (dimensionless) dissipation 
rate is connected to this problem, in the sense that correlation (alignment) 
of the different vector fields could influence the energy transfer 
across the scales \cite{Pouquet78,Biskamp93,Dallas13b}, and thus possibly the amount of 
energy that is eventually dissipated at the small scales. 

For neutral fluids it has been known for a long time that
the dimensionless dissipation rate in forced and freely decaying homogeneous
isotropic turbulence tends to a constant with increasing Reynolds number. The first 
evidence for this was reported by Batchelor \cite{Batchelor53} in 1953, while the 
experimental results reviewed by Sreenivasan in 1984 \cite{Sreenivasan84}, and subsequent
experimental and numerical work by many groups, 
established the now well-known characteristic curve of the dimensionless dissipation 
rate against Reynolds number: see
\cite{Sreenivasan98, McComb14a, McComb14c, Vassilicos15} and references therein.
For statistically steady isotropic turbulence, the theoretical explanation of this 
curve was recently found to be connected to the energy balance equation for forced turbulent
flows \cite{McComb14c}, where the asymptote describes the maximal inertial transfer 
flux in the limit of infinite Reynolds number.   
 
For freely decaying MHD, recent results suggest that the temporal 
maximum of the total dissipation tends to a constant value with increasing Reynolds number.    
 The first evidence for this behavior in MHD
  was put forward in 2009 by Mininni and Pouquet \cite{Mininni09}
 using results from direct numerical simulations (DNSs) of isotropic MHD turbulence. 
 The temporal maximum of the total dissipation rate $\vep(t)$ became independent of Reynolds number
 at a Taylor-scale Reynolds number $\Rl$ (measured at the peak of $\vep(t)$) of about 200.
 
 Dallas and Alexakis \cite{Dallas14b} measured the dimensionless dissipation
 rate $\Ceps$ from DNS data, where $\vep$ was non-dimensionalized with respect to the initial
 values of the rms velocity $U(t)$ and the integral length scale $L(t)$ 
 (here defined with respect to the total energy),
 for random initial fields with strong correlations between the velocity field
 and the current density.
 The authors compared data with Ref.~\cite{Mininni09},
 and again it was found that $\Ceps \to const.$ with increasing Reynolds number. Interestingly
 the approach to the asymptote was slower than for the data of Ref.~\cite{Mininni09}.
 
In this Letter we propose a model for the Reynolds number dependence 
of the dimensionless dissipation rate derived from the energy balance 
equation for MHD turbulence in terms of Els\"{a}sser fields \cite{Politano98}, 
which predicts nonuniversal values of the dimensionless
dissipation rate in the infinite Reynolds number limit.
In order to compare the predictions of the model against data, we carried out
a series of DNSs of decaying isotropic MHD turbulence. 
Firstly we explain the derivation of the model equation, then proceed to 
a description of our numerical simulations and subsequently compare the model 
to DNS results. We conclude with a discussion of the results and suggestions for 
further research.

The equations describing incompressible decaying MHD flows are
\begin{align}
\label{eq:momentum}
\partial_t \vec{u}&= - \frac{1}{\rho}\nabla P -(\vec{u}\cdot \nabla)\vec{u}
 + \frac{1}{\rho}(\nabla \times \vec{b}) \times \vec{b} + \nu \Delta \vec{u}  \ , \\
\label{eq:induction}
\partial_t \vec{b}&= (\vec{b}\cdot \nabla)\vec{u}-(\vec{u}\cdot \nabla)\vec{b} + \eta \Delta \vec{b}\ , \\
\label{eq:incompr}
&\nabla \cdot \vec{u} = 0 \ \ \mbox{and} \ \  \nabla \cdot \vec{b} = 0 \ ,  
\end{align}
where $\vec{u}$ denotes the velocity field, $\vec{b}$ the magnetic 
induction expressed in Alfv\'{e}n units, $\nu$ the kinematic viscosity, 
$\eta$ the resistivity, $P$ the pressure and $\rho=1$ the density.
For simplicity and in order to compare to results in the literature we 
consider the case of unit magnetic Prandtl number, that is $Pm=\nu / \eta =1$. 

For freely decaying MHD turbulence the decay rate of the total energy
$\vep_D=-\partial_t E_{tot}$ equals the total dissipation rate $\vep$, 
and the time evolution of the total energy is governed by the energy balance
equation of MHD turbulence in real space, which is derived from the MHD equations
\eqref{eq:momentum}-\eqref{eq:incompr}. 
This suggests that the energy balance equation can be used 
in order to derive the Reynolds number dependence of the total dissipation rate.

Since we are interested in the total dissipation 
$\varepsilon = \varepsilon_{mag} + \varepsilon_{kin}$, 
where
$\vep_{mag} = 2 \eta \int_0^{\infty} dk \ k^2 E_{mag}(k)$ ,
and
$\vep_{kin} = 2 \nu \int_0^{\infty} dk \ k^2 E_{kin}(k)$ ($E_{mag}(k)$ and $E_{kin}(k)$ denoting
magnetic and kinetic energy spectra),
are the magnetic and kinetic dissipation rates, respectively,
 we could take two approaches, either formulating the energy
balance in terms of the primary fields $\vec{u}$ and $\vec{b}$ or 
in terms of the Els\"{a}sser fields $\vec{z}^{\pm}=\vec{u} \pm \vec{b}$. Since 
\beq
\label{eq:eps_Hc_z}
\partial_t \langle |\vec{z}^\pm|^2\rangle = 2\partial_t E_{tot} \pm 2\partial_t H_c \ ,
\eeq
where $H_c=\langle \vec{u} \cdot \vec{b}\rangle$ is the cross helicity,
we can describe the total dissipation either by the energy balance
equations for $\langle |\vec{z}^\pm|^2\rangle$ \cite{Politano98} or by 
the sum of the energy balance equations for 
$E_{mag}(t)=\int_0^{\infty} dk \ E_{mag}(k)$ and $E_{kin}(t)=\int_0^{\infty} dk \ E_{kin}(k)$ \cite{Chandrasekhar51,Podesta08}.

This, however, is not the case if we are interested in the {\em dimensionless} 
dissipation rate.  
Unlike in hydrodynamics, there are several choices of scales 
with which to non-dimensionalize $\vep(t)$ and thus with respect to which 
to define an MHD analogue to the Taylor surrogate expression \cite{Batchelor53, McComb14a}.
For example $U$ and $L$ could be used, 
or the rms $\vec{b}$ field $B$ and $L$ or $U$ and $L_{kin}$ etc., 
or scales defined with respect to $\vec{z}^{\pm}$. The physical 
interpretation is different for the different scaling quantities. Since the total 
dissipation must equal the total flux of energy passed through the scales by the kinetic and 
magnetic energy transfer terms, a scaling with $U$ will be appropriate only for 
hydrodynamic transfer as this transfer term scales as $U^3/L_{kin}$. 
All other transfer terms include 
$\vec{b}$ and $\vec{u}$ and thus should be scaled accordingly. This also precludes
the most straightforward generalization of the Taylor surrogate,  
which would be a scaling of $\vep$ with $L$ and $\sqrt{U^2 + B^2}$. A hydrodynamic 
transfer term would then be scaled partly with magnetic quantities, while the 
appropriate scaling should 
only involve kinetic quantities.

Instead we propose to define the dimensionless dissipation rate for MHD turbulence with respect to 
the Els\"{a}sser variables 
\beq
\Ceps = \frac{\Ceps^+ + \Ceps^-}{2} \equiv \frac{1}{2} \left ( \frac{\vep L_+}{{z^+}^2 z^-} + \frac{\vep L_-}{{z^-}^2 z^+} \right ) \ ,
\label{eq:ceps_defn}
\eeq  
where $L_{\pm}=(3\pi \int_0^{\infty} dk \ k^{-1} \langle |\vec{z}^\pm|^2 \rangle)/(4\int_0^{\infty} dk \ \langle |\vec{z}^\pm|^2\rangle)$ 
are the integral scales defined with respect to $\vec{z}^\pm$, and 
$z^{\pm}$ denote the rms values of $\vec{z}^{\pm}$ \footnote{
The scaling is ill-defined for the (measure zero) cases $\vec{u} = \pm \vec{b}$,
which correspond to exact solutions to the MHD equations where
the nonlinear terms vanish. Thus no turbulent transfer is possible, 
and these cases are not amenable to an analysis
which assumes nonzero energy transfer
\cite{Politano98}.}. 

Using this 
definition we can now consistently non-dimensionalize the evolution equations of 
$\langle |\vec{z}^\pm|^2\rangle$. For conciseness we outline the arguments for the
$\langle |\vec{z}^+|^2\rangle$ case, since the  
$\langle |\vec{z}^-|^2\rangle$ case proceeds analogously 
\cite{SM}.
    
Following \cite{Politano98} the energy balance for $\langle |\vec{z}^+|^2\rangle$ reads for the case $Pm=1$
\begin{align}
\label{eq:PP}
-\frac{1}{2}\partial_t \langle |\vec{z}^+|^2 \rangle&= 
-\frac{3}{4} \partial_t B_{LL}^{++} - \frac{\partial_r}{r^4} 
\left(\frac{3r^4}{2}C^{+-+}_{LL,L} \right) \nonumber \\
&+\frac{3(\nu + \eta)}{2r^4} \partial_r \left(r^4 \partial_rB_{LL}^{++} \right) \ ,
\end{align}
where 
$C^{+-+}_{LL,L}(r)$ and $B_{LL}^{++}(r)$
are the longitudinal third-order correlation function and the 
second-order longitudinal structure function of the Els\"{a}sser fields, respectively. 
The definitions of the these functions can be found in the Supplemental Material \cite{SM}. 
Using \eqref{eq:eps_Hc_z} one can express the LHS of \eqref{eq:PP}
in terms of $\vep(t)$ and $\partial_t H_c$. 

If we now introduce the nondimensional variable $\sigma=r/L_+$ \cite{Wan12} and 
 nondimensionalize equation \eqref{eq:PP} with respect to $z^{\pm}$ and 
$L_+$ as proposed in the definition of $\Ceps$ in eq.~\eqref{eq:ceps_defn}, we obtain 
\begin{align}
\label{eq:evol_z_scaled}
\Ceps^+
&= -\frac{\partial_{\sigma}}{\sigma^4} \left(\frac{3\sigma^4C^{+-+}_{LL,L}}{2{z^+}^2 z^-}\right) -\frac{L_+}{{z^+}^2 z^-}\partial_t \frac{3B_{LL}^{++}}{4} 
 \nonumber \\ 
& +\frac{L_+}{{z^+}^2 z^-}\partial_t H_{c} +\frac{\nu+\eta}{L_+{z}^-} \frac{3\partial_{\sigma}}{2\sigma^4} 
\left(\sigma^4 \partial_{\sigma}\frac{B_{LL}^{++}}{{z^+}^2} \right) \ .
\end{align}
In this way we arrive at a consistent scaling for each
transfer term in \eqref{eq:PP} with the appropriate
quantity, as the function $C^{+-+}_{LL,L}(r)$ 
scales with ${z^+}^2 z^-$. 

Since the inverse of the coefficient in front of the dissipative term 
is similar to a Reynolds number, we introduce the 
generalized large-scale Reynolds number 
\beq
R_-=\frac{2z^-L_+}{\nu+\eta} \ , 
\label{eq:gen_Rey}
\eeq
hence \eqref{eq:evol_z_scaled} suggests a dependence of $\Ceps^+$ on $1/R_-$. 
However, the structure and correlation functions and the cross helicity flux
also depend on Reynolds number.

For conciseness we introduce dimensionless versions of all terms 
present on the RHS of 
\eqref{eq:evol_z_scaled}, such that 
\begin{align}
C^{+-+}_{LL,L}(r,t)&= {z^+}^2z^-g^{+-+}(\sigma,t) , \  \\
B_{LL}^{++}(r,t)&= {z^+}^2h_2^{++}(\sigma,t) , \ \\
\partial_t B_{LL}^{++}(r,t)&= \frac{({z^+})^2z^-}{L_+}F^+(\sigma,t) , \ \\
\partial_t H_c(t)&= \frac{({z^+})^2z^-}{L_+}G^+(t) , \ 
\end{align}  
which leads to a dimensionless version of the $\langle |\vec{z}^+|^2\rangle$ 
energy balance equation for freely decaying MHD turbulence 
\begin{align}
\label{eq:evol_z_nondim}
\Ceps^+ =\frac{\vep L_+}{{z^+}^2z^-}
&= -\frac{\partial_{\sigma}}{\sigma^4} \left(\frac{3\sigma^4}{2}g^{+-+}\right) 
-\frac{3}{4} F^+ +G^+ \nonumber \\
&+\frac{3}{R_-} \frac{\partial_{\sigma}}{\sigma^4} \left(\sigma^4 \partial_{\sigma}h_2^{++} \right) \ .
\end{align}
After non-dimensionalization the highest derivative in the differential equation 
is multiplied with the small parameter $1/R_-$, suggesting that this can 
be viewed as a singular perturbation problem \cite{Wasow65}; and thus we consider 
asymptotic expansions of the dimensionless functions in
inverse powers of $R_-$ \cite{Lundgren02,McComb14c}. 

The formal asymptotic series of a generic function $f$ (used for conciseness in place of 
the functions on the RHS of \eqref{eq:evol_z_scaled}) 
up to second order in $1/R_-$ reads
\begin{equation}
\label{eq:asymp_F}
f= f_{0}+ \frac{1}{R_-}f_{1} 
+ \frac{1}{R_-^2}f_{2} + O(R_-^{-3}) \ .  
\end{equation}  

After substitution of the expansions into 
\eqref{eq:evol_z_nondim} and following the same steps for 
the evolution equation for $\langle |\vec{z}^-|^2\rangle$,
 we arrive at model equations for $\Ceps^+$ and $\Ceps^-$
\beq
\Ceps^\pm = \Cinf^\pm + \frac{C^\pm}{R_{\mp}} + \frac{D^\pm}{R_{\mp}^2} + O(R_{\mp}^{-3}) \ ,
 \label{eq:model+}
\eeq 
up to third order in $1/R_{\mp}$, where we defined the coefficients 
$\Cinf^\pm$, $C^\pm$ and $D^\pm$ 
\beq
\label{eq:cinf+}
\Cinf^\pm = -\frac{\partial_{\sigma}}{\sigma^4} \left(\frac{3\sigma^4}{2}g_0^{\pm\mp\pm}\right)
-\frac{3}{4} F_0^\pm  \pm  G_0^\pm \ , \\
\eeq
\begin{align}
\label{eq:c+}
C^\pm &=\frac{3\partial_{\sigma}}{\sigma^4} \left[ \sigma^4 \left( \partial_{\sigma} 
h_{2,0}^{\pm\pm} - \frac{g_1^{\pm\mp\pm}}{2} \right)\right] 
-\frac{3}{4} F_1^\pm \pm G_1^\pm  \ , \\ 
\label{eq:d+}
D^\pm &=\frac{3\partial_{\sigma}}{\sigma^4} \left[ \sigma^4 \left( \partial_{\sigma} 
h_{2,1}^{\pm\pm} - \frac{g_2^{\pm\mp\pm}}{2} \right)\right] 
-\frac{3}{4} F_2^\pm \pm G_2^\pm \ , 
\end{align}
in order to write \eqref{eq:evol_z_nondim} in a more concise way.
Using $R_+ = (L_-/L_+)(z^+/z^-)R_-$ to define
\beq
C= \frac{1}{2}\left ( C^+ + \frac{L_-}{L_+} \frac{z^+}{z^-} C^-\right ) \ ,
\eeq 
($D$ is defined analoguously), finally one obtains for the dimensionless 
dissipation rate $\Ceps$
\beq
\Ceps = \Cinf + \frac{C}{R_-} + \frac{D}{R_-^2} + O(R_-^{-3}) \ .
 \label{eq:model}
\eeq

Since the time dependence of the various quantities in this problem has been 
suppressed for conciseness, we stress 
that \eqref{eq:model} is time dependent, including the Reynolds 
number $R_-$. A normalization using initial values of $z^\pm$ and $L^\pm$ would have resulted in
a dependence of $\Ceps(t)$ on initial values of $R_-$, which only describe the  
initial conditions and not the evolved flow for which $\Ceps$ is measured.

At the peak of $\vep(t)$ the additional terms $F_0^\pm$ should in fact 
vanish for constant flux of cross helicity 
(that is, $\partial_t^2 H_c = 0$), 
since in the infinite Reynolds number limit the 
second-order structure function will have its inertial range form
at all scales. By self-similarity the spatial and temporal dependences of e.g.~$B_{LL}^{++}$ 
should be separable in the inertial range, that is $B_{LL}^{++}(r,t) \sim (\vep^+(t)r)^{\alpha}$ 
for some value $\alpha$, and 
$\partial_t B_{LL}^{++} \sim \alpha\vep^+(t)^{\alpha-1}\partial_t \vep^+ r^\alpha$.
At the peak of dissipation $\partial_t \vep^+|_{t_{peak}} = 
\partial_t\vep|_{t_{peak}} -\partial_t^2 H_c = \partial_t\vep|_{t_{peak}} =0$, 
and we obtain $F_0^+(t_{peak})=0$.
As the terms $G_0^\pm$ which describe the flux of 
cross helicity in the infinite Reynolds number limit, cancel
the corresponding contribution from the transfer terms \cite{SM},
the asymptotes $\Cinf^\pm$ describe the flux of total energy
provided the model \eqref{eq:model+} is applied at $t_{peak}$.

Due to selective decay, that is the faster decay of the total energy 
compared to $H_c$ and $H_{mag}$ \cite{Biskamp93}, one could perhaps expect $\partial_t H_c$ to 
be small compared to $\vep$ in the infinite Reynolds number limit in most situations. 
In this case we obtain $G_0^{\pm} \simeq 0$ and
\beq
\Cinf^\pm(t_{peak}) = -\frac{\partial_{\sigma}}{\sigma^4} \left(\frac{3\sigma^4}{2}g_0^{\pm\mp\pm} \right) \ ,
\eeq
which recovers the inertial-range scaling results of Ref.~\cite{Politano98} and reduces 
to Kolmogorov's 4/5th law for $\vec{b}=0$. 

Since $\Cinf$ is a measure of the flux of total energy across different scales in 
the inertial range, differences for the value of this asymptote should be 
expected for systems with different initial values for the ideal invariants 
$H_c$ and magnetic helicity $H_{mag}=\langle \vec{a} \cdot \vec{b}\rangle$, where
$\vec{a}$ is the vector potential $\vec{b}=\nabla \times \vec{a}$. 
In case of $H_{mag} \neq 0$, the value of $\Cinf$ should
be {\em less} than for $H_{mag}=0$ due to a more pronounced reverse
energy transfer in the helical case \cite{Pouquet78}\footnote{Reverse transfer of magnetic 
energy has recently been discovered in nonhelical 3D MHD turbulence
\cite{Brandenburg15,Zrake14}}, the result of which is {\em less} forward transfer and 
thus a smaller value of the flux of total energy. 
For $H_c \neq 0$ we expect $\Cinf$ to be smaller than for $H_c =0$, 
since alignment of $\vec{u}$ and $\vec{b}$ weakens the coupling
of the two fields in the induction equation, which leads to less transfer
of magnetic energy across different scales and presumably also less transfer
of kinetic to magnetic energy. 
In short, one should expect nonuniversal values of $\Cinf$. 

Before we compare the model equation with DNS data and 
address this question of nonuniversality numerically, we briefly outline 
our numerical method. 
Equations \eqref{eq:momentum}-\eqref{eq:incompr} are solved numerically
in a periodic box of length $L_{box}=2 \pi$ 
using a fully de-aliased pseudospectral MHD code \cite{Yoffe12, Berera14a}.
All simulations satisfy $k_{max}\eta_{mag,kin} \geqslant 1$, where
$\eta_{mag,kin}$ are the magnetic and kinetic Kolmogorov scales, respectively.
We do not impose a background magnetic field, and both the initial magnetic
and velocity fields are random Gaussian with zero mean, with initial magnetic 
and kinetic energy spectra of the form
$E_{mag,kin}(k)\sim k^4\exp(-k^2/(2k_0)^2)$,
where $k_0 \geqslant 5$ and further simulation details are specified in 
Table 1 of \cite{SM}.
The initial relative magnetic helicity is $\rho_{mag}(k)=kH_{mag}(k)/2E_{mag}(k)=1$
for all runs of series H and zero for the runs labelled
NH. The initial relative cross helicity was 
$\rho_c(0)=H_c(0)/(|\vec{u}(0)||\vec{b}(0)|)=0$ 
for runs of the H and NH series and $\rho_c(0)=0.6$ 
for series CH06H and CH06NH, while initial 
magnetic and kinetic energies were in equipartition.
All spectral quantities have been shell- and ensemble-averaged, with ensemble sizes
restricted by computational resources to up to 10 runs per ensemble. 
The total dissipation rate $\vep$ was measured at its maximum.

\begin{figure}[!t]
 \begin{center}
 \includegraphics[width=\columnwidth]{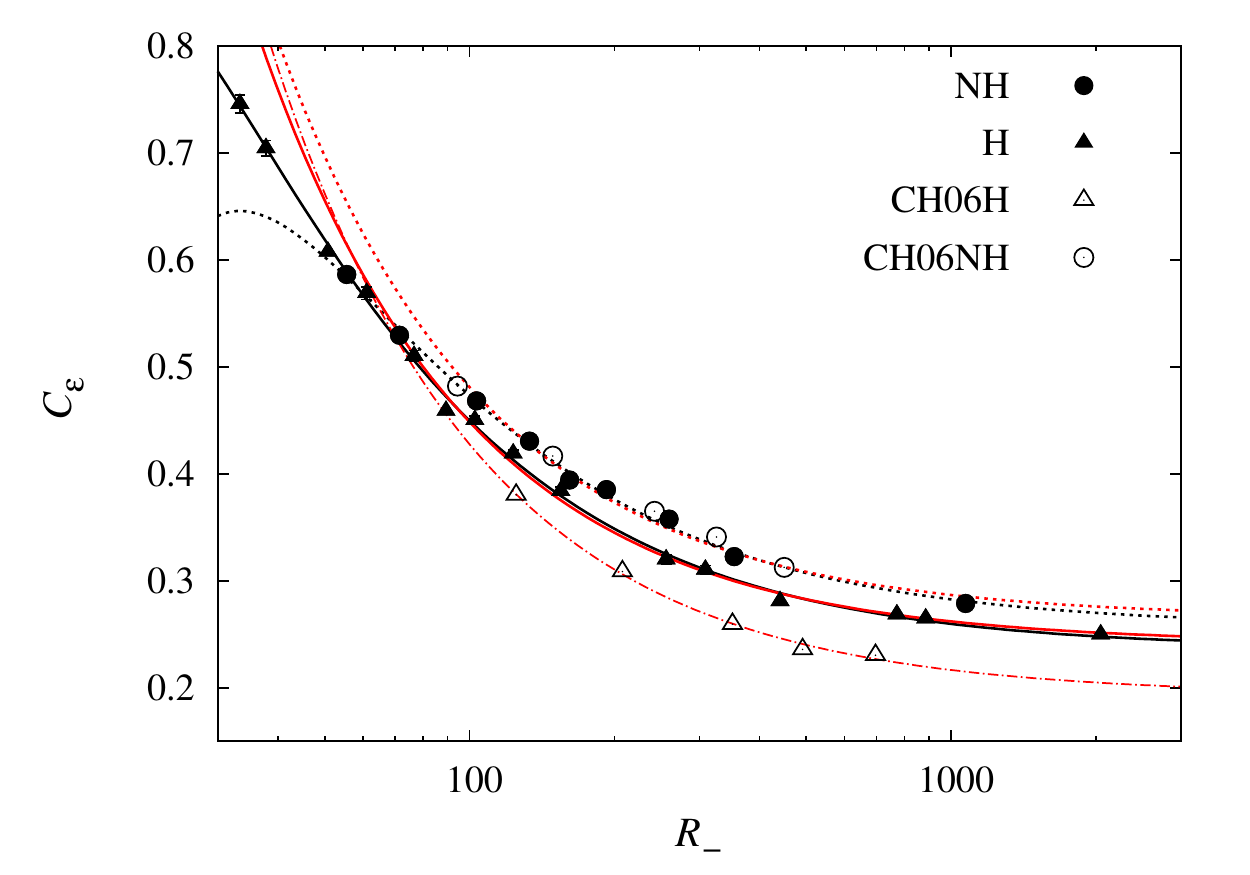}
 \caption{(Color online) The solid and dotted and dash-dotted lines show \eqref{eq:model} fitted to
  helical, non-helical and cross-helical DNS data, respectively. The red (grey) lines refer to fits using
  the first-order model equation, the black lines use the model equation up to second order in 
  $1/R_-$.
  As can be seen, the respective asymptotes differ for the data sets.
 } 
 \label{fig:comp_nonhel}
 \end{center}
\end{figure}

Figure \ref{fig:comp_nonhel} shows fits of the model equation to DNS 
data for datasets that differ in the initial 
value of $H_{mag}$ and $H_c$. As can be seen, the model fits the data very well.
For the series H runs and for $R_- > 70$ 
it is sufficient to consider terms of first order in $R_-$, 
while for the series NH the first-order approximation is valid for $R_- > 100$.
The cross-helical CH06H runs gave 
consistently lower values of $\Ceps$ compared to the series H runs, while
little difference was observed between series CH06NH and NH. 
The asymptotes were $\Cinf=0.241 \pm 0.008$ 
for the H series, $\Cinf=0.265 \pm 0.013$ 
for the NH series, $\Cinf=0.193 \pm 0.006$ for the CH06H series
and $\Cinf=0.268 \pm 0.005$ for the CH06NH series.

As predicted by the qualitative theoretical arguments outlined before, 
the measurements show that the asymptote calculated from the nonhelical runs 
is larger than for the helical case, 
as can be seen in Fig.~\ref{fig:comp_nonhel}. 
The asymptotes of the series H and NH do not lie within one standard 
error of one another. 
Simulations carried out with $H_c \neq 0$ suggest little difference
in $\Ceps$ for magnetic fields with initially zero magnetic helicity. For initially
helical magnetic fields $\Ceps$ is further quenched if $H_c \neq 0$.
In view of nonuniversality, an even larger variance
of $\Cinf$ can be expected once other parameters such as external forcing, 
plasma $\beta$, $Pm$, etc., are taken into account. Here we have restricted ourselves
to nonuniversality caused by different values of vector field correlations.   

In summary, a definition for the dimensionless dissipation rate 
$\Ceps$ for MHD turbulence has been proposed, 
where $\vep$ was non-dimensionalized with respect to the Els\"{a}sser fields 
instead of the rms velocity. For this definition of $\Ceps$ and the case of 
unit Prandtl number we derived a model for the dependence of $\Ceps$ on a 
generalized Reynolds number $R_-$.  
The model predicts that $\Ceps \to const$ with 
increasing $R_-$, in analogy to hydrodynamics, and the asymptote is a measure of the total
energy transfer flux. 
The model was compared to DNS data for datasets which differ in their initial values of 
magnetic and cross helicities. At moderate to high $R_-$, we found good agreement 
to data with the model only using terms up to first order in $1/R_-$. 
However, at low $R_-$ terms of second order in $R_-$ cannot be neglected,
\moritz{in fact these terms improve the fit specifically at low $R_-$. 
This is expected from adding another term in the expansion 
and thus provides further justification of the validity of eq.~\eqref{eq:model}.}    

As predicted, the values of the respective asymptotes from the datasets differ, 
suggesting a dependence of $\Ceps$ on different values of the helicities, 
and thus a connection to the question of universality in MHD turbulence. 
This presents an interesting point for further research 
concerning the influence of other vector field correlations on the dissipation rate. 
Other questions concern the generalization of this approach to more general 
MHD systems such as flows with magnetic Prandtl numbers $Pm \neq 1$, 
\moritz{compressive fluctuations}, 
and to the presence of a background magnetic field, 
as well as to turbulent systems 
where the flow carries other quantities such as temperature or 
pollutants; and also the application 
to decaying hydrodynamic turbulence \cite{Vassilicos15}. 
In the most general case in plasmas there will be a mean 
magnetic field, which leads to spectral anisotropy \moritz{and the breakdown of the
conservation of magnetic helicity \cite{Matthaeus82a}} and thus might 
introduce several difficulties to be overcome when generalizing this method, 
as the spectral flux will then depend on the direction of the mean 
field \cite{Wan09,Wan12} \moritz{and a more generalized description 
and role for the magnetic helicity would be needed}.

Our model shows that different degrees of correlation in a turbulent 
plasma control the amount of energy that can effectively be transferred into 
the smallest scales. It could have several possible applications, e.~g.~ 
for heating rates in the solar wind, especially as
high values of the cross helicity inhibit such transfer to some extent.
For situations where one is interested in sustaining a magnetic field over long times, 
thus trying to minimize dissipative effects, one could estimate from \eqref{eq:cinf+}-\eqref{eq:c+}  
what type of correlations produce not only a low asymptotic value of the dissipation rate 
but also a fast approach to this asymptote. 
This would have relevance to cosmological and astrophysical \cite{Sorriso07}
magnetic fields as well as terrestial plasmas,  
such as in a tokamak reactor.
Our results suggest that in cosmology, where a topical problem is 
the origin of large-scale magnetic fields, it is not only a nonzero value of magnetic 
helicity, but perhaps also the parameter range of other correlations such as the cross
and kinetic helicities, that facilitate the presence of long-time magnetic fields. 
Moreover, this raises questions about the possible generation
mechanisms for cosmological magnetic fields leading to different correlations between 
the vector fields such that they can sustain long evolution times.

The authors thank Samuel Yoffe for providing his
hydrodynamic pseudospectral code for further development
\moritz{and the Referees and the Editor for insightful comments}.
This work has made use of the resources provided by ARCHER
({\tt http://www.archer.ac.uk}), made available through the Edinburgh
Compute and Data Facility (ECDF, 
{\tt http://www.ecdf.ed.ac.uk}).
A.~B. is supported by STFC, M.~F.~L. and M.~E.~M. are funded by 
\moritz{the UK Engineering and Physical Sciences 
Research Council (EP/K503034/1 and EP/M506515/1)}.

\bibliography{magn,wdm}

\end{document}

\title{Supplemental material for \\
``Nonuniversality and finite dissipation in decaying magnetohydrodynamic turbulence"}

\author{M.~F. Linkmann}
\email[]{m.linkmann@ed.ac.uk}
\affiliation{SUPA, School of Physics and Astronomy, University of Edinburgh, Peter Guthrie Tait Road, EH9 3FD, UK}

\author{A. Berera}
\email[]{ab@ph.ed.ac.uk}
\affiliation{SUPA, School of Physics and Astronomy, University of Edinburgh, Peter Guthrie Tait Road, EH9 3FD, UK}

\author{W.~D. McComb}
\affiliation{SUPA, School of Physics and Astronomy, University of Edinburgh, Peter Guthrie Tait Road, EH9 3FD, UK}

\author{M.~E. McKay}
\affiliation{SUPA, School of Physics and Astronomy, University of Edinburgh, Peter Guthrie Tait Road, EH9 3FD, UK}

\begin{abstract}
In this Supplemental Material we include definitions of the structure and 
correlation functions used in the Letter. For simulations with zero cross 
helicity we give numerical verification that the two model equations for 
$\Ceps^+$ and $\Ceps^-$ in fact lead to identical results, as expected.   
Moreover, we provide a table of simulation details with information on the 
resolution of our simulations, Reynolds numbers and measured values of the 
dimensionless dissipation rate.  
\end{abstract}

\maketitle
\section{Definitions of the structure and correlation functions}
In order to keep this material self-consistent, we include here the definitions
of the Els\"{a}sser structure and correlation functions used in the Letter. 
The third-order correlation and structure functions $C^{\pm\pm\mp}_{LL,L}(r)$, 
$C^{\pm\mp\pm}_{LL,L}(r)$ and $B_{LL,L}^{\pm\mp\pm}(r)$ and the second-order structure functions 
$B_{LL}^{\pm\pm}(r)$ are defined as follows:
\begin{align}
\label{eq:strfct1}
B_{LL,L}^{\pm\mp\pm}(r)&= \langle (\delta_L z^\pm(r))^2 \delta_L z^\mp(r)\rangle , \  \\
\label{eq:strfct2}
C^{\pm\pm\mp}_{LL,L}(r)&= \langle z_L^\pm(\vec{x}) z_L^\pm(\vec{x}) z_L^\mp(\vec{x} + \vec{r}) \rangle , \  \\
\label{eq:strfct3}
C^{\pm\mp\pm}_{LL,L}(r)&= \langle z_L^\pm(\vec{x}) z_L^\mp(\vec{x}) z_L^\pm(\vec{x} + \vec{r}) \rangle , \  \\
\label{eq:strfct4}
B_{LL}^{\pm\pm}(r)&= \langle (\delta_L z^\pm(r))^2 \rangle , \ 
\end{align}
 where
$v_L = \vec{v}\cdot\vec{r}/r$ denotes the longitudinal component of a vector field $\vec{v}$,
that is its component parallel to the displacement vector $\vec{r}$, and  
\beq
\delta_L v(r) = [\vec{v}(\vec{x}+\vec{r})-\vec{v}(\vec{x})]\cdot \frac{\vec{r}}{r} \ ,
\eeq
its longitudinal increment. 

The transfer term in the energy balance equations for $\vec{z}^\pm$ are written 
in the Letter in terms of the function $C^{\pm\pm\mp}_{LL,L}(r)$ for reasons of conciseness.
It is common in the literature to use the relation
\beq
C_{LL,L}^{\pm\mp\pm}=\frac{1}{4}\left (B_{LL,L}^{\pm\mp\pm}-2C_{LL,L}^{\pm\pm\mp} \right) \ ,
\eeq
in order to express the transfer terms through the functions $B_{LL,L}^{\pm\mp\pm}(r)$ and 
$C^{\pm\pm\mp}_{LL,L}(r)$ instead \cite{Politano98}.

\section{Validation for different scalings for zero cross helicity}
The definition of $\Ceps$ as in eq.~(5) of the Letter was
proposed in order to arrive at a consistent scaling for the
transfer terms in the energy balance equations for $\vec{z}^\pm$,
since the functions $C^{\pm\mp\pm}_{LL,L}(r)$
which describe the transfer terms scale as $(z^\pm)^2z^\mp$.
This led to separate model equations for $\Ceps^+$ and $\Ceps^-$.
In the Letter, we outlined the derivation of the model equation for $\Ceps^+$ 
from rescaling the evolution equation for $\langle |\vec{z}^+|^2\rangle$ 
leading to 
\beq
\Ceps^+ = \Cinf^+ + \frac{C^+}{R_{z^-}} + \frac{D^+}{R_{z^-}^2} 
+ O(R_{z^-}^{-3}) \ .
 \label{eq:model+}
\eeq
In the same way a model equation for 
\beq
\Ceps^- = \frac{\vep L_{z^-}}{{z^-}^2 z^+} \ ,
\label{eq:ceps_defn}
\eeq
is derived through nondimensionalizing the energy balance equation 
for $\vec{z}^-$. The flux terms in this equation are given in terms 
of the function $C^{-+-}_{LL,L}(r)$ 
(or equivalently $C^{--+}_{LL,L}(r)$ and $B_{LL,L}^{-+-}(r)$), and the 
dissipative term in terms of $B_{LL}^{--}(r)$.

For zero cross helicity one should expect $\Ceps^+ = \Ceps^-$, since 
\beq
\langle |\vec{z}^+|^2\rangle = 2E_{tot} + 2H_c = 2E_{tot} \ \
\mbox{and }
\langle |\vec{z}^-|^2\rangle = 2E_{tot} - 2H_c = 2E_{tot} \ .
\eeq 
Therefore all quantities defined with respect to the rms fields $z^+$ and 
$z^-$ should be the same, and thus for zero cross helicity it is possible to 
use, $\Ceps^+$ or $\Ceps^-$ representatively for $\Ceps$. 
  
\begin{figure}[tbp]
 \begin{center}
 \includegraphics[width=0.45\textwidth]{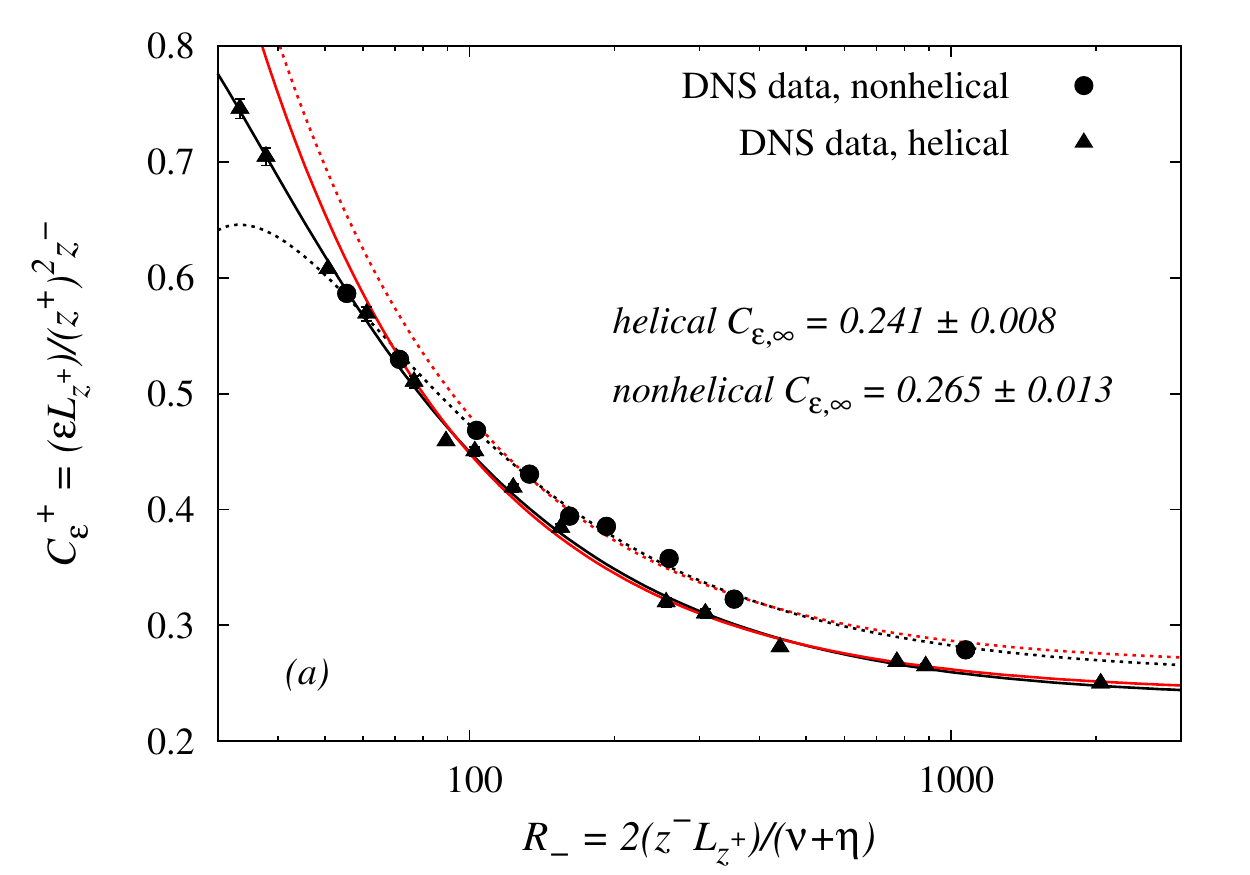}
 \includegraphics[width=0.45\textwidth]{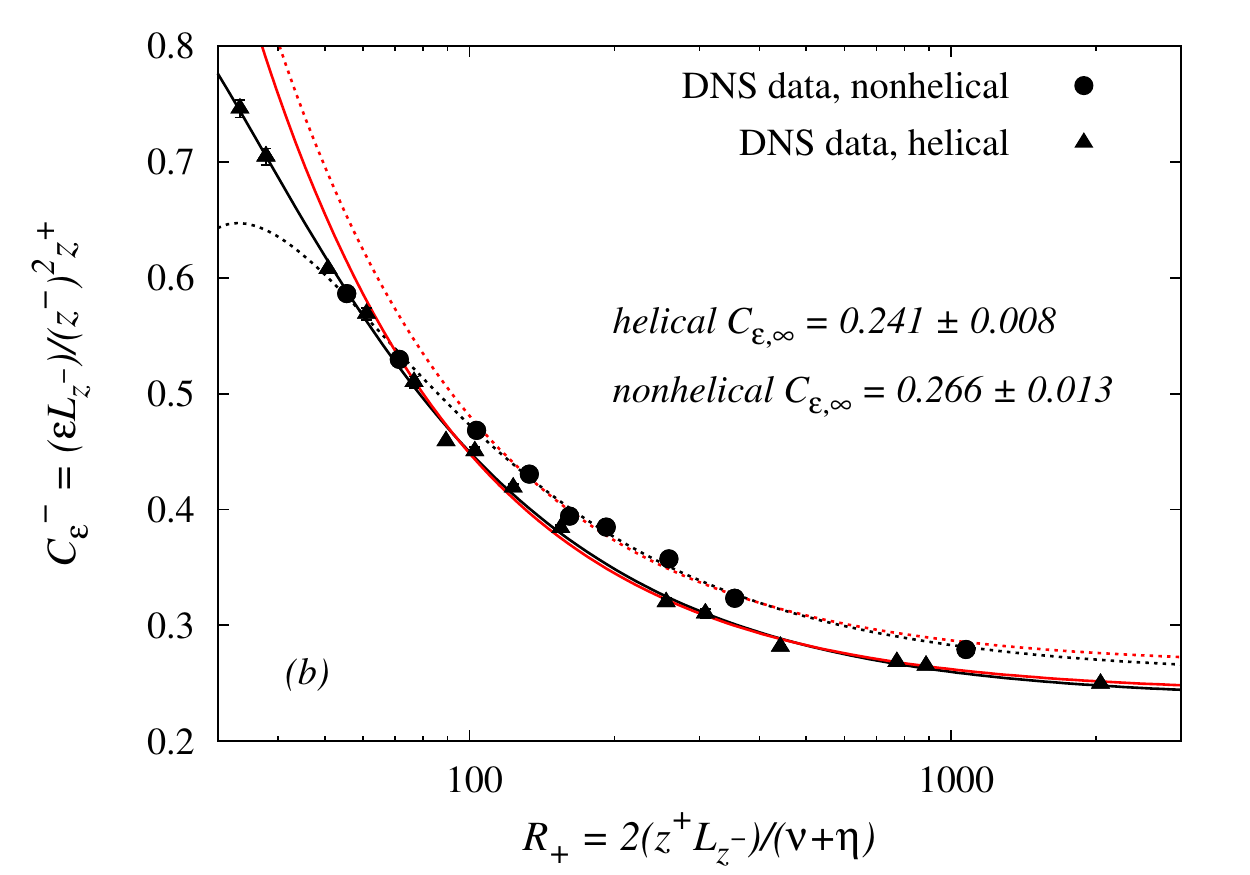}
 \caption{The solid and dotted lines show the model equations 
 fitted to helical and non-helical DNS data, respectively. Figure (a) shows
 $\Ceps^+$ and Fig.~(b) $\Ceps^-$. In both figures, the red lines refer to 
fits using the first-order model equations, the black lines use the model 
equations up to second order in $1/R_{z^\pm}$. By comparing Figs.~(a) and (b) 
it can clearly be seen that both scalings give the same results.
 }
 \label{fig:validation}
 \end{center}
\end{figure}

Figure \ref{fig:validation} shows separate fits of the model equations for 
$\Ceps^+$ and $\Ceps^-$ (obtained by the two nondimensionalizations) 
to DNS data for zero cross helicity. 
The measured asymptotes are identical
and the curves are practically indistinguishable from each other and 
from those shown in Fig.~(1) of the Letter, which was obtained by fitting eq.~(20)
of the Letter to the data. This provides numerical verification for the representative use
of $\Ceps^+$ (or $\Ceps^-$) to describe $\Ceps$.
 
\begin{figure}[tbp]
 \begin{center}
 \includegraphics[width=0.45\textwidth]{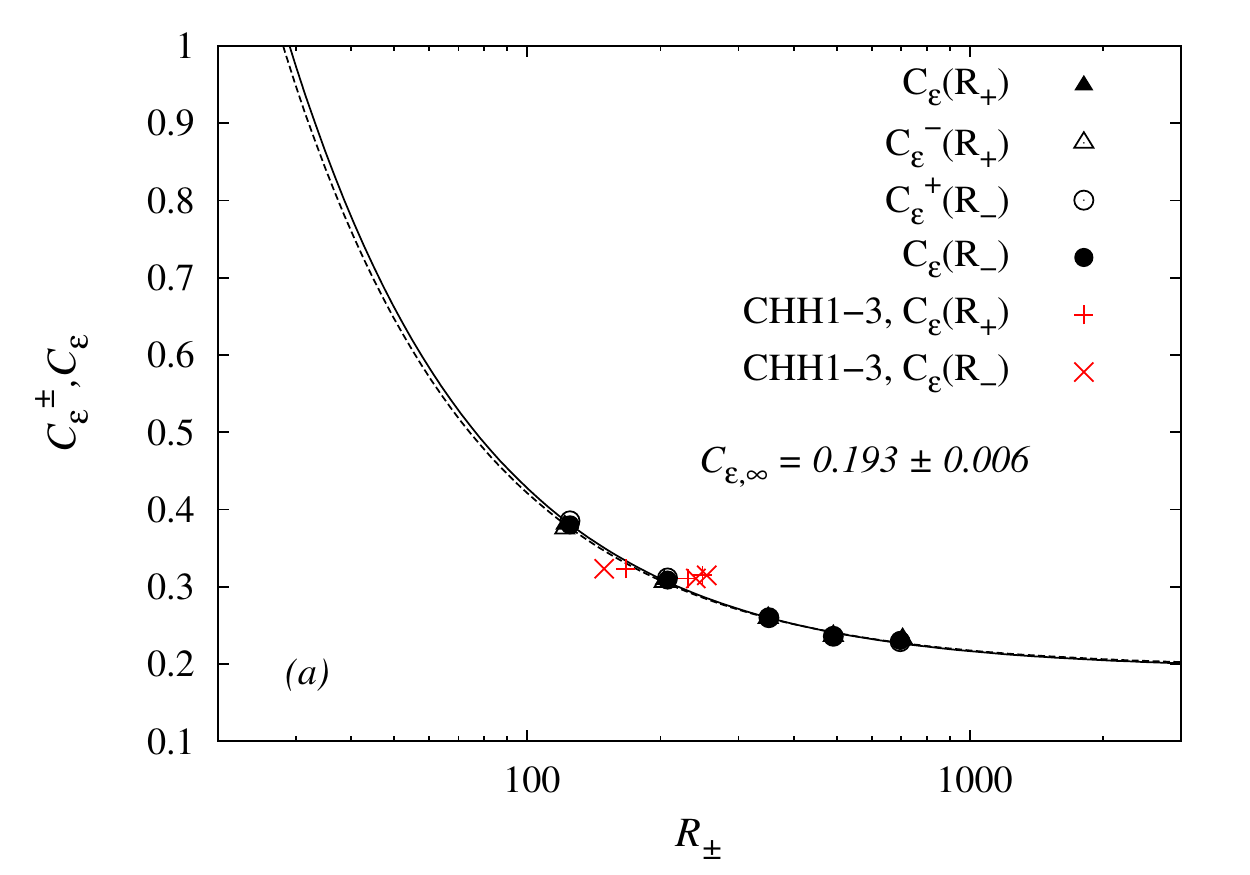}
 \includegraphics[width=0.45\textwidth]{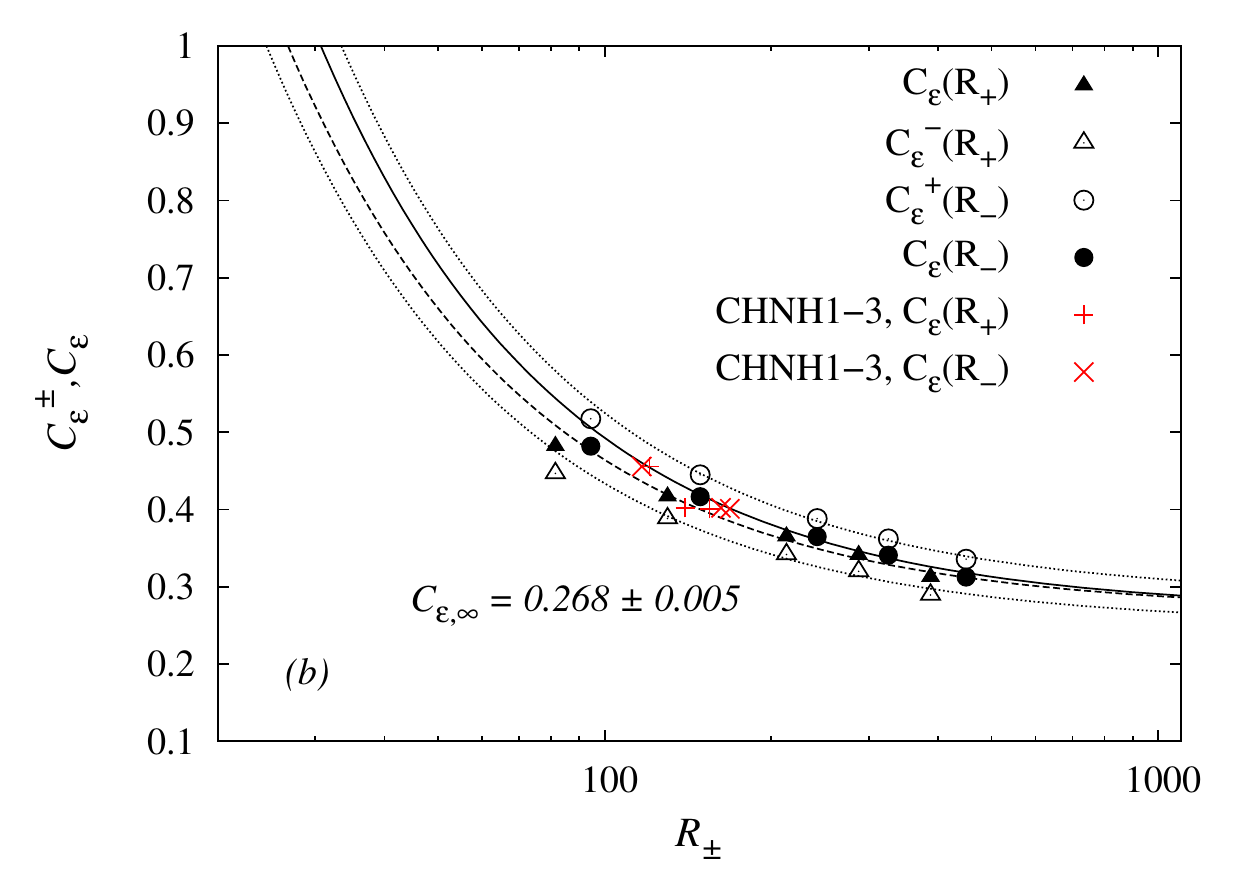}
 \caption{The solid lines show the model equations
 for $\Ceps$, $\Ceps^+$ and $\Ceps^-$ fitted to data for 
(a) series CH06H and (b) series CH06NH with initial relative cross helicity $\rho_c(0)=0.6$. 
 The red crosses refer to additional runs with $0.2 \leqslant \rho_c(0) \leqslant 0.8$.
 }
 \label{fig:valid_crosshel}
 \end{center}
\end{figure}

However, for nonzero cross helicity the difference in the rms values 
$z^+$ and $z^-$ precludes the use of $\Ceps^+$ (or $\Ceps^-$) to represent the 
dimensionless dissipation rate 
$\Ceps$, and both $\Ceps^+$ and $\Ceps^-$ have to be taken into account 
separately, which is reflected in the definition of the dimensionless 
dissipation rate in eq.~(5) of the Letter. Figure \ref{fig:valid_crosshel}
shows data for nonzero cross helicity $H_c$, that is runs of series CH06H and CH06NH, as well 
as the additional series CHH and CHNH.
Interestingly, very little difference between $\Ceps^+$ and $\Ceps^-$ is observed for series 
CH06H. For series CH06NH, note that the asymptote $\Cinf$ is the same, no matter
if extrapolated using $R_-$ or $R_+$.

\section{Simulation specifications}
%%%%%%%%%%%%%%%%%%simulation specs%%%%%%%%%%%%%%%%%%%%%%%%
 \begin{table}[!h]
 \begin{center}
 \begin{ruledtabular}
 \begin{tabular}{ccccccccccccc}
   Run id & $N^3$  & $k_{max}\eta_{mag}$ & $R_{-}$ & $R_L$ & $\Rl$ &  $\eta $  & $k_{0}$ & \#  & $\Ceps$ 
  & $\sigma$ & $\rho_c(0)$ \\
  \hline
   H1 & $128^3$ & 1.30 & 33.37 & 25.28 & 14.87 & $0.009$ & 5 & 10 & 0.756 & 0.008 & 0\\
  H2 & $256^3$ & 2.42 & 37.77 & 27.81 & 15.85 & $0.008$ &  5 & 10 & 0.704  & 0.007 & 0\\
%  H3 & $256^3$ & 2.18 & & 36.88 &  & $0.007$  & 5 & 10 &  &  \\
  H3 & $512^3$ & 1.38 & 50.81 & 35.08 & 18.55 & $0.002$  & 15 & 10  & 0.608 & 0.001 & 0 \\
% H3a & $256^3$ & 2.08 & 50.23 & 34.33 & 17.86 & $0.006$  & 5 & 10 & 0.584 &0.044\\
  H4 & $256^3$ & 1.80 & 61.14 & 40.63 & 20.34 & 0.005  &  5 & 10 & 0.569 & 0.006 & 0 \\
  H5 & $256^3$ & 1.59 & 76.72 & 48.73 & 23.11 & 0.004 &  5 & 10 & 0.510 & 0.005 & 0\\
  H6 & $1024^3$ & 1.38 & 89.32 &55.51 & 25.76 & $0.00075$  & 23 & 10 & 0.4589 & 0.0003 & 0\\
  H7 & $256^3$ & 1.29 & 102.53 & 60.65 & 26.91 & $0.003$ & 5 & 10 & 0.450 & 0.004 & 0\\
  H8 & $512^3$ & 2.33 & 123.17 & 69.40 & 29.67 & $0.0025$ & 5 & 10 & 0.419 & 0.003 & 0\\
  H9 & $512^3$ & 2.01 & 154.67 & 83.06 & 33.84 & $0.002$ & 5 & 10 & 0.384 & 0.003 & 0\\
  H10 & $512^3$ & 1.45 & 255.89 & 123.97 & 45.21 & $0.0012$ & 5 & 10 & 0.320 & 0.004 &0\\
  H11 & $528^3$ & 1.31 & 308.69 & 143.71 & 50.18 & $0.001$ & 5 & 10 & 0.310 & 0.004& 0\\
  H12 & $1024^3$ & 2.03 & 441.25 & 194.38 & 61.39 & $0.0007$  & 5 & 5 & 0.281 & 0.002 & 0\\
  H13 & $1032^3$ & 1.38 & 771.34 & 309.08 & 82.97 & $0.0004$  & 5 & 5 & 0.268 & 0.001 & 0\\
  H14 & $1024^3$ & 1.24 & 885.05 & 358.72 & 88.76 & $0.00035$  & 5 & 5 & 0.265 & 0.002 & 0\\
  H15 & $2048^3$ & 1.35 & 2042.52 & 724.71 & 136.25 & $0.00015$  & 5 & 1 & 0.250 & - & 0\\
  \hline
  CH06H1 & $512^3$ & 2.17 & 124.89 & 108.81 & 49.88 & $0.002$ & 5 & 1 & 0.380 & - & 0.6 \\
  CH06H2 & $512^3$ & 1.57 & 207.61 & 171.87 & 68.57 & $0.0012$ & 5 & 5 & 0.309 & 0.002 & 0.6 \\
  CH06H3 & $1024^3$ & 2.21 & 351.52 & 277.21 & 95.31 & $0.0007$ & 5 & 1 & 0.260 & - & 0.6 \\
  CH06H4 & $1024^3$ & 1.76 & 491.50 & 380.70 & 116.85 & $0.0005$ & 5 & 1 & 0.236 & - & 0.6 \\
  CH06H5 & $1024^3$ & 1.37 & 696.19 & 523.08 & 132.48 & $0.00035$ & 5 & 1 & 0.231 & - & 0.6 \\
  CHH1 & $512^3$ & 1.46 & 254.64 & 129.83 & 47.55 & $0.0012$ & 5 & 5 & 0.315  & 0.002  & 0.2 \\
  CHH2 & $512^3$ & 1.50 & 240.35 & 147.68 & 55.25 & $0.0012$ & 5 & 5 & 0.311 & 0.003 & 0.4 \\
  CHH3 & $512^3$ & 1.74 & 149.28 & 195.27 & 90.60 & $0.0012$ & 5 & 5 & 0.323  & 0.03 & 0.8 \\
  \hline
%  NH1 & $128^3$ & 1.29 & & 28.69  & & $0.009$ & 5 & 10 & 50 & \\
 % NH2 & $256^3$ & 2.42 & 32.27 & & $0.008$  & 5 & 10 & 50 & \\
  NH1 & $256^3$ & 1.51 & 55.57 & 53.89 & 25.57 & $0.004$  & 5 & 10 & 0.587 &0.005 & 0 \\
  NH2 & $256^3$ & 1.26 & 71.51 & 68.60 & 30.11 & $0.003$ & 5 & 10 & 0.530 &0.004 & 0\\
  NH3 & $512^3$ & 1.86 & 103.41 & 96.69 & 37.68 & $0.002$  & 5 & 10 & 0.468 &0.004 & 0\\
  NH4 & $512^3$ & 1.51 & 133.14 & 122.51 & 43.94 & $0.0015$  & 5 & 10 & 0.431 &0.004 & 0\\
  NH5 & $512^3$ & 1.29 & 161.35 & 151.76 & 50.73 & $0.0012$  & 5 & 10 & 0.394 &0.004 & 0\\
  NH6 & $1024^3$ & 2.28 & 192.40 & 168.28 & 54.44 & $0.001$ & 5 & 5 & 0.358 & 0.002 & 0\\
  NH7 & $1024^3$ & 1.76 & 259.58 & 232.10 & 65.42 & $0.0007$ & 5 & 5 & 0.358 & 0.002 & 0 \\
  NH8 & $1024^3$ & 1.40 & 354.30 & 301.71 & 76.73 & $0.0005$  & 5 & 5 & 0.323 & 0.002 & 0\\
  NH9 & $2048^3$ & 1.15 & 1071.44 & 823.58 & 134.73 & $0.00015$  & 5 & 1 & 0.279 & - & 0\\
  \hline
  CH06NH1 & $512^3$ & 2.02 & 94.39 & 113.02 & 49.29 & $0.002$ & 5 & 1 & 0.482 & - & 0.6 \\
  CH06NH2 & $512^3$ & 1.41 & 148.86 & 174.61 & 65.48 & $0.0012$ & 5 & 5 & 0.417  & 0.003 & 0.6 \\
  CH06NH3 & $1024^3$ & 1.93 & 242.06 & 272.85 & 87.50 & $0.0007$ & 5 & 1 & 0.365 & - & 0.6 \\
  CH06NH4 & $1024^3$ & 1.52 & 325.62 & 365.54 & 104.45 & $0.0005$ & 5 & 1 & 0.341  & - & 0.6 \\
  CH06NH5 & $1024^3$ & 1.16 & 450.01 & 515.23 & 127.09 & $0.00035$ & 5 & 1 & 0.313 & - & 0.6 \\
  CHNH1 & $512^3$ & 1.31 & 168.31 & 152.84 & 51.40 & $0.0012$ & 5 & 5 & 0.401 & 0.003 & 0.2 \\
  CHNH2 & $512^3$ & 1.34 & 162.27 & 157.48 & 55.52 & $0.0012$ & 5 & 5 & 0.402 & 0.006 & 0.4 \\
  CHNH3 & $512^3$ & 1.58 & 116.71 & 204.78 & 87.52 & $0.0012$ & 5 & 5 & 0.456 & 0.002 & 0.8 \\
  \end{tabular}
  \end{ruledtabular}
  \end{center}
 \caption{Specifications of simulations. % \cite{data}.
%\footnote{The data is publicly available at {\tt http://datashare.is.ed.ac.uk}}. 
H refers to an initially helical
 magnetic field, NH to an initially non-helical magnetic field.
 $R_L$ denotes the integral-scale Reynolds number,
 $R_{\lambda}$ the Taylor-scale Reynolds number,
 $R_{-}$ the generalized Reynolds number as in given in eq.~(8) of the Letter, $\eta$ the
 magnetic resistivity, $k_0$ the peak wavenumber of the initial energy spectra,
 $k_{max}$ the largest resolved wavenumber, $\eta_{mag}$ the Kolmogrov microscale
 associated with the magnetic field at the peak ot total dissipation,
 \# the ensemble size, $\Ceps$ the dimensionless total dissipation rate defined in eq.~(5) of the Letter, $\sigma$ the
 standard error on $\Ceps$ and $\rho_c(0)$ the initial relative cross helicity. All Reynolds numbers are measured
 at the peak of total dissipation. 
 }
 \label{tbl:simulations}
 \end{table}

\section{Energy and cross-helicity fluxes}

%The transfer terms in the energy balance equations for 
%$\langle |\vec{z}^\pm|^2\rangle$ are given by spatial derivatives of the
%third order correlation function
%\beq
%C^{\pm\mp\pm}_{LL,L}(r) = \langle \vec{z}_L^\pm(\vec{x}) \vec{z}_L^\mp(\vec{x}) \vec{z}_L^\pm(\vec{x} + \vec{r}) \rangle  \ .
%\eeq

In the Letter it is pointed out that the asymptotes $\Ceps^\pm$ describe 
the total energy flux, as the contribution of the cross-helicity flux to the
Els\"{a}sser flux is cancelled by the respective terms $G_0^\pm$. Here we provide some 
further details.   
The asymptotes $\Cinf^\pm$ were given in the Letter as 
\beq
\label{eq:cinf+}
\Cinf^\pm = -\frac{\partial_{\sigma}}{\sigma^4} \left(\frac{3\sigma^4}{2}g_0^{\pm\mp\pm}\right)
-\frac{3}{4} F_0^\pm  \pm  G_0^\pm \ ,
\eeq
which reduced at the peak of dissipation to 
\beq
\label{eq:cinf+_tp}
\Cinf^\pm = -\frac{\partial_{\sigma}}{\sigma^4} \left(\frac{3\sigma^4}{2}g_0^{\pm\mp\pm}\right)  \pm  G_0^\pm \ .
\eeq
The dimensional version of this equation is 
\beq
\label{eq:cinf+_tp_dim}
\vep = -\frac{\partial_r}{r^4} \left(\frac{3r^4}{2}C_{LL,L}^{\pm\mp\pm}(r)\right)  \pm  \partial_t H_c \ ,
\eeq
where we assume that the function $C_{LL,L}^{\pm\mp\pm}$ has its inertial range form corresponding to $g_0^{\pm\mp\pm}$.
The function $C_{LL,L}^{\pm\mp\pm}$ can be written in terms of $B_{LL,L}^{\pm\mp\pm}$ and $C_{LL,L}^{\pm\pm\mp}$
\beq
C_{LL,L}^{\pm\mp\pm}=\frac{1}{4}\left (B_{LL,L}^{\pm\mp\pm}-2C_{LL,L}^{\pm\pm\mp} \right) 
=\frac{1}{4}\left ( \langle (\delta_L z^\pm(r))^2 \delta_L z^\mp(r)\rangle 
-2\langle z_L^\pm(\vec{x}) z_L^\pm(\vec{x}) z_L^\mp(\vec{x} + \vec{r}) \rangle \right ) \ ,
\eeq
which can be written in terms of the primary fields $\vec{u}$ and $\vec{b}$ as
\begin{align}
\label{eq:CLL_ub}
C_{LL,L}^{\pm\mp\pm}&=
\frac{1}{4}\left ( \langle (\delta_L z^\pm(r))^2 \delta_L z^\mp(r)\rangle 
-2\langle z_L^\pm(\vec{x}) z_L^\pm(\vec{x}) z_L^\mp(\vec{x} + \vec{r}) \rangle \right ) \nonumber \\
&= \frac{1}{4}\frac{2}{3}\langle(\delta_L u(r))^3 - 6  b_L(\vec{x})^2 u_L(\vec{x}+\vec{r}) \rangle
\mp \frac{1}{4} \frac{2}{3}\langle(\delta_L (b(r))^3 - 6  u_L(\vec{x})^2 b_L(\vec{x}+\vec{r}) \rangle \ ,
\end{align}
see e.g.~\cite{Politano98}, 
%By comparison to the evolution equations of the total energy 
%\begin{align}
%\label{eq:etot_evol}
%-\partial_t E_{tot}(t)=&-\partial_t \langle (\delta_L b(r))^2 +(\delta_L u(r))^2  \rangle
% +\frac{\partial_r}{r^4} \left(\frac{r^4}{4} \left[\langle(\delta_L u(r))^3 - 6  b_L(\vec{x})^2 u_L(\vec{x}+\vec{r}) \rangle \right] \right) \nonumber \\
%&+ \frac{\nu+\eta}{r^4} \partial_r \left (r^4 \partial_r \langle (\delta_L b(r))^2 +(\delta_L u(r))^2  \rangle \right) \ ,
%\end{align}
%and the cross helicity
%\begin{align}
%-\partial_t H_c =&
%-\partial_t \langle (\delta_L b(r))^2 +(\delta_L u(r))^2  \rangle
% +\frac{\partial_r}{r^4} \left(\frac{r^4}{4} \left[\langle(\delta_L b(r))^3 - 6  u_L(\vec{x})^2 b_L(\vec{x}+\vec{r}) \rangle \right] \right) \nonumber \\
%&+ \frac{\nu+\eta}{r^4} \partial_r \left (r^4 \partial_r \langle (\delta_L b(r))^2 +(\delta_L u(r))^2  \rangle \right) \ ,
%\end{align}
and the two terms on the
last line of \eqref{eq:CLL_ub} are related to the flux 
terms in the evolution equations of the total energy and 
the cross helicity \cite{Politano98}. 

Now going back to \eqref{eq:cinf+_tp_dim}, we can write in the inertial range
\begin{align}
\label{eq:cinf+_tp2}
\vep &= -\frac{\partial_r}{r^4} \left(\frac{3r^4}{2}C_{LL,L}^{\pm\mp\pm}(r)\right)  \pm  \partial_t H_c \nonumber \\
&=-\frac{\partial_r}{r^4} \left(\frac{r^4}{4} \left [ \langle(\delta_L u(r))^3 - 6  b_L(\vec{x})^2 u_L(\vec{x}+\vec{r}) \rangle 
\mp  \langle(\delta_L (b(r))^3 - 6  u_L(\vec{x})^2 b_L(\vec{x}+\vec{r}) \rangle \right ] \right) \pm \partial_t H_c \nonumber \\
&= \vep_T \pm \vep_C \pm \partial_t H_c = \vep_T \ ,
\end{align}
where $\vep_T$ is the flux of total energy and $\vep_C$ 
the cross-helicity flux, which must equal $-\partial_t H_c$ in the inertial range
for freely decaying MHD turbulence, thus we see that the 
contribution from the third-order correlator $C_{LL,L}^{\pm\mp\pm}$ 
resulting in $\vep_C$ is cancelled by $\partial_t H_c$, or, 
after nondimensionalization, the cross helicity flux
$\vep_C L_{\pm}/[(z^\pm)^2z^-]$ is cancelled by $G_0^\pm$.   

\bibliography{magn,wdm}